# A Brand-new Research Method of Neuroendocrine System

Sheng-Rong Zou, Zhong-Wei Guo, Yu-Jing Peng,

Ta Zhou，Chang-Gui Gu, Da-Ren He*

Yangzhou University, Yangzhou 225009, P. R. China

**Abstract:**

In this paper, we present the empirical investigation results on the neuroendocrine system by bipartite graphs. This neuroendocrine network can describe the structural characteristic of neuroendocrine system. The act degree distribution and cumulate act degree distribution show so-called "shifted power law" (SPL) function forms. And the similarity of neuroendocrine network is $s = 0.14$. Our results may be used in the research of the medical treatment of neuroendocrine diseases.

**Keywords**：Complex networks, Bipartite graphs, Act degree distribution, Mediator, Bioinformatics.

E-mail:guozhongwei@126.com   telephone: 0086 13773417646   fax:0086 514 87887937

# 1. Introduction

A plethora of evidence, accumulated mainly during the first half of the 20th century, indicates that the endocrine and nervous systems integrate and regulate different body functions. Rich interconnections take place between neural, endocrine systems, which may constitute a neural–endocrine functional complex (Boris Mravec, 2006).The neuroendocrine system comprises a network of specialized neurons and endocrine cells that regulate metabolism, reproduction, blood pressure, mood, stress, thermoregulation, sleep, body fluid and electrolyte homeostasis (Gabriele Di Comite, 2007). For the precise coordination of systemic functions, the nervous system uses a variety of peripherally and centrally localized receptors, which transmit information from internal and external environments to the central nervous system. Tight interconnections between the nerve and endocrine systems provide a base for monitoring and consequent modulation system functions by the brain and vice versa (Boris Mravec, 2006).

For decades, neuroendocrine researchers have employed standard statistical methods to uncover the relationships between variables and constructs. However, in complex neuroendocrine systems, the prevalence of complex and outliers is to be expected. Under such circumstances, the use of standard statistical methods becomes unreliable and, correspondingly, results in degraded predictions of the relationships within the neuroendocrine systems (John Grznar, 2007). We try to describe the neuroendocrine systems so as to provide more accurate direction for clinical treatment and pharmacy.

Some authors have modeled the neuroendocrine system, with a variety of approaches and areas of emphasis (Catherine R., 2007. Kirsten Labudda, 2007. Magali Malacombe, 2006. Robert H. Bonneau, 2007. Stéphane Culine, 2007). But many essential features of this complex system are still not understood. Many of the systems can be effectively modeled by networks, in which the units of the systems are modeled as nodes, and the interactions among the units are modeled concisely as edges (with or without weights) between the nodes. This represents a new framework for building models of complex systems (M. E. J. Newman, 2002. R. Albert, 2002). In this paper we suggest a neuroendocrine network with bipartite graphs theory (Petter Holme, 2003).

The paper is organized as follows. We will present the method in the second section, in which we will describe the neuroendocrine system with the method. In section three, we will introduce and study the system empirically. In the last part, we make a conclusion.

**Our method**

The recent years have witnessed an upsurge in the study of complex systems (Pei-Pei Zhang, 2006). We note, especially, that some scientists used bipartite graph as a powerful tool (R. Albert, 2002). In this graph one type of vertices is "actors" taking part in some activities, organizations or events. The other type of vertices is the activity, organization or event named "acts" (Hui Chang, 2007). Many systems are naturally modeled as bipartite networks: Biochemical networks can be described by vertices representing chemical substances separated by edges representing chemical reactions (H. Jeong, 2000). Such studies have been very active recently, due to their complex structure, to the ubiquity of such bipartite networks in complex systems, and to the large databases available (M. E. J. Newman, 2002).

According to the complex network theory, the neuroendocrine system can be represented by bipartite graphs. Usually in bipartite graphs only the edges between different types of nodes are considered as shown in Fig. 1 where the larger black circles labeled by 1, 2, 3 denote acts, the smaller void circles labeled by A, B, C, D denote actors. The line segments denote edges, which represents actors' participation in acts (Yong-Zhou Chen, 2007). In the neuroendocrine system, cells can be seen as a certain nodes, the mediators secreted by cells as other nodes. We regard every kind of cells as an act, mediators an actor, the secretion relation between cells and mediators sides and the mediators secreted by manifold cells bridges linking these acts. All of these constitute our neuroendocrine network.

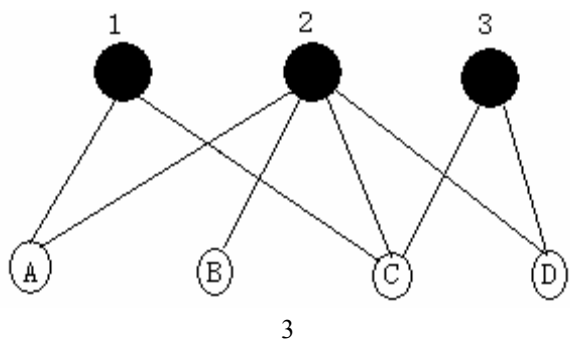

Fig.1. The projection of a schematic neuroendocrine bipartite graph.

**Empirical results of the act degree and similarity**

We empirically research neuroendocrine network. We collect data through COPE database and construct the network．Fig.2 shows the act degree distribution, P(h), of the neuroendocrine network, and Fig.3 shows the accumulative act degree distribution, P(h≥hc), of the neuroendocrine network. The two distributions shown in Fig. 2 and Fig. 3 can be well described with SPL (Hui Chang, 2007) functions.

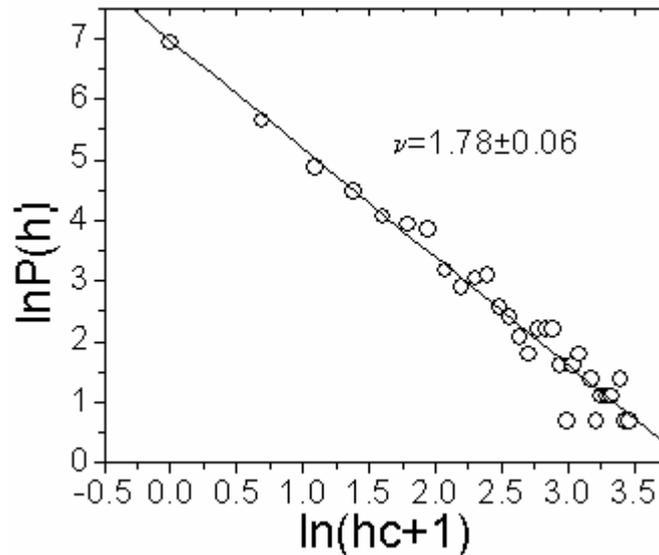

Fig. 2. The empirical results on the act degree distribution of neuroendocrine network. The solid lines represent the least square fitting of the data.

The act degree and accumulative act degree are two important geometric properties of complex networks. The act degree stands for the times that an actor takes part in acts. In neuroendocrine network, the act degree stands for the number of the cells that secretes a single mediator, in which bFGF(basic fibroblast growth factor) is the largest node act degree. It is an important mitogenic cytokine, followed by TGF-beta, IL-6, IL1-beta, VEGF, IGF-1and so on .They are mediators that can promote the growth and differentiation of a variety of cells. They are secreted by kinds of cells and become the largest bridge among acts in the network, for the reason that the most basic condition of body survival is the multiply of cells and when the body is injured, it needs the rapid growth of the corresponding cells to secrete some special mediators to promote wound healing and tissue repair. Whereas, the mediators with little node act degree have certain unique functions, and they are from a certain category or categories of secretory cells, such as adiponectin, adipsin, CCL12, dopamine, epinephrine. These mediators can only be secreted in a particular environment to control the physical feeling, behavior and so on. Therefore our conclusions are reasonable. The average act degree of neuroendocrine network is h = 3.01, It means each mediator is secreted by three cells on an average. Fig. 2 shows the act degree distribution of the neuroendocrine network. The distribution can be well described using SPL, which illuminates that in the process of network formation the dynamics mechanism includes both preferential principles, with a probability p, and randomly selected factors, with the probability 1-p. And that means the cells secrete mediators helpful for their growth and differentiation. As some mediators can be secreted by most cells, they are preferred. Also there are too many specific factors lacking interconnections (the types of diseases, conditions and the states of the patients), the other mediators are selected randomly. Because there are too many factors which are not associated, it is random selection.

Accumulative act degree (multiple edges are counted) means the times that an actor takes part in acts. Fig. 3 shows the accumulative act degree distribution (with multiple edges counted) of the neuroendocrine network.

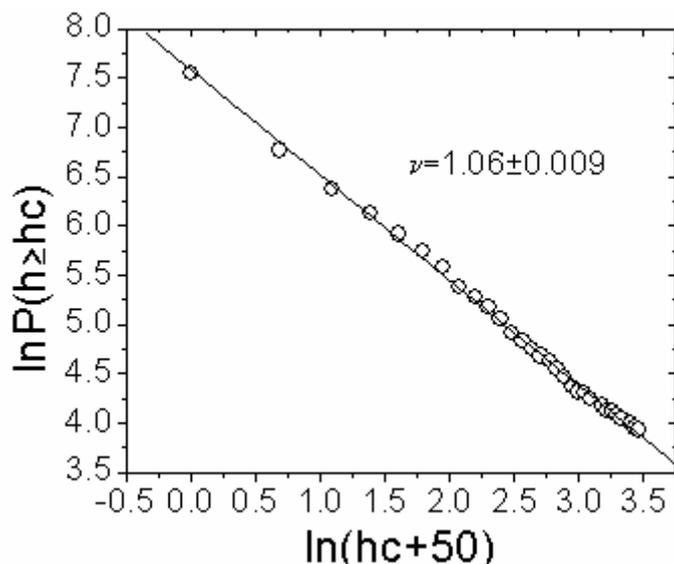

Fig. 3. The empirical results on the accumulative act degree distribution (multiple edges are counted) of neuroendocrine network. The solid lines represent the least square fitting of the data.

We also have calculated the similarity of neuroendocrine network, namely, the average probability of secreting the same mediators by all the neuroendocrine cells. It can be written as:

$$s = \frac{\sum_i \frac{\sum_{i'} \frac{\sum_j e_{ij} e_{i'j}}{\sqrt{k_i k_{i'}}}}{N}}{N}$$

Where N is the total number of cells, $k_i$ is the degree of cell i, namely the number of mediators connected by cells, $e_{ij}$ is the connecting side situation of cell i and mediator. When the sides are connected, $e_{ij}$ equals to 1, when not connected, $e_{ij}$ equals to 0. If the similarity is 1, it illuminates each cell functions exactly the same. If the similarity is 0, it illuminates each cell functions differently.

The similarity of neuroendocrine network is r = 0.14. This shows that there is little probability of secreting the same mediators by different cells; most of the mediators secreted are different. There is little similarity in cells and the functions are varied from each other. These are similar to our intuitive understanding. For example, astrocytes play an active role in brain function, and possibly even direct the activities of neurons. It prompts glutamate to produce special molecules that nourish neurons, known as trophic factors, while pancreatic acinar cells secrete enzymes which take part in digestion, and hormonal regulation of blood glucose levels.

**Conclusion and discussion**

The contributions of our paper propose a new approach for research neuroendocrine system, which were not previously treated in this way. In our method there are two types of basic elements. One of them can be addressed as acts, the other called actors.

The investigations propose some conclusions. Firstly, act degree distribution, accumulative act degree distribution of neuroendocrine network show SPL function forms, which can continuously vary from an ideal power law to an ideal exponential decay. Secondly, we have found several very important mediators with our method, such as bFGF, TGF-beta, IL-6, IL1-beta, VEGF, IGF-1, CCL2, TNF-alpha, IL-8, M-CSF. They are relatively highly act degree. It reveals these mediators are in neuroendo-crine system to maintain bodily healthiness, emotional stabilization and endocrine harmony. Thirdly, In the neuroendocrine system we can find some immune cytokines (for example, IL-1, IL-2, GM-CSF and so on) and them receptors. So that we can deeply understand the relation between neuroendocrine system and the immune system, and further understand how the neuroendocrine system influence and regulate the immune system. Finally, we can use similarity to measure the similar degree of function between cells.

These conclusions are important, however, it is possible that we have not discovered the most important properties, because there are not enough data. Even though, the obtained empirical statistical properties already show some interesting conclusions, for example, the act degree of mediator should be very interesting for a lot of scientists. Therefore, further research is necessary to illuminate the physiological mechanism of the neuroendocrine system.

**Acknowledgements**

The research is supported by National Natural Science Foundation of China under the grant No. 70671089 and 10635040.